\documentclass[aps,amsmath,amssymb,reprint,superscriptaddress]{revtex4-1}

\usepackage[usenames]{color}
\usepackage{colortbl}
\usepackage{amsmath}
\usepackage{graphicx}
\usepackage[right]{lineno}
\usepackage{dcolumn}

\begin{document}

%\preprint{}

%\title{}


%\author{}
%\thanks{}
%\affiliation{}

\title{Two components of donor-acceptor recombination in compensated semiconductors.
Analytical model of spectra in presence of electrostatic fluctuations}

\author{N.A. Bogoslovskiy}
\author{P.V. Petrov}
\email{pavel.petrov@gmail.com}
\author{Yu.L. Iv\'anov}
\thanks{Deceased 18 June 2018}
\author{K.D. Tsendin}
\thanks{Deceased 4 April 2018}
\author{N.S. Averkiev}
\affiliation{Ioffe Institute, St. Petersburg, Russian Federation}


%\date{\today}

%\linenumbers

\begin{abstract}
We report numerical and analytical studies of the donor-acceptor recombination in compensated semiconductors.
Our calculations take into account random electric fields of charged impurities which are
important in non zero compensation case. We show that the donor-acceptor optical spectrum
can be described as a sum of two components: monomolecular and bimolecular.
In the low compensation limit we develop two analytical models for both
types of the recombination. Also our numerical simulation predicts that these two components
of the photoluminescence spectra can be resolved under certain experimental conditions.
\end{abstract}

%\pacs{}
%\keywords{}

\maketitle

\section{Introduction}
The donor-acceptor (DA) optical transition is due to a tunnel recombination
between distant impurity pairs \cite{PhysRev.140.A202}.
If radii of both impurity states are comparable with the lattice constant
then the spectrum of DA recombination appears as a set of narrow
lines which corresponds to discreet positions of impurities in a crystal
lattice. If one of the impurity states (donor, for instance) is shallow and can be
described with the hydrogen-like wave function with radius $a_{b}$ then DA recombination corresponds
to a broad spectrum. The probability $p$ of the transition
exponentially decreases with inter-impurity distance $r$. The photon
energy $\hbar\omega$ increases due to Coulomb interaction $e^2/\varepsilon r$ between ionized
donor and acceptor in the final state:
\begin{equation}
\label{p}
p=p_0 \exp{\left(-\frac{2r}{a_b}\right)};
\end{equation}
\begin{equation}
\label{hw}
\hbar \omega = E_g - E_d - E_a + \frac{e^2}{\varepsilon r}.
\end{equation}
Here $E_g$ is the bandgap of the material, $E_d$ and $E_a$ --- donor and
acceptor binding energies, $p_0$ is a part of matrix element which does not depend on a distance.
For short we use the denotation $E = \hbar\omega - (E_g - E_d - E_a)$ which we call the
transition energy in the rest of the paper. 

It is easy to derive the spectral dependence of the transition probability $P(E)$
\cite{PhysRev.140.A202, osipov1976theory, Bogoslovskiy2016}.
The probability to find a donor at a distance $r$ from an arbitrary acceptor
equals $N_d 4 \pi r^2 dr$, where $N_d$ is the concentration of donors. Calculating $dr=({dr}/{dE})dE$ from equation (\ref{hw})
and using $E_d = {e^2}/({2 \varepsilon a_b})$ we obtain:
\begin{equation}
\label{FO}
P(E)dE = 32\pi p_0 N_d a_b^3 \left(\frac{E_d}{E}\right)^{4} \exp{\left(-\frac{4E_d}{E}\right)} \frac{dE}{E_d},
\end{equation}
This expression is normalized per one photo-excited acceptor.

Using the same approach one can describe kinetic
properties of DA recombination. After an optical pumping pulse
nearest DA pairs recombine faster producing photons with higher
energies. Distant pairs recombine slowly therefore DA luminescence line
shifts with time to long wavelengths \cite{PhysRev.140.A202, pssb19680250202, levanyuk}.

Such an approach is acceptable in the case of infinitely low
compensation. In the presence of compensation the DA luminescence
spectrum is broadened by random electric fields of ionized impurities. An
impact of this fields is a non-trivial question of solid state physic.
Besides fundamental aspects the problem of DA spectrum is interesting for
modern technology applications. Nowadays new semiconductor compounds
constantly come into view and attract a lot of attention as perspective materials
for optoelectronics and photovoltaics. Any new semiconductor material 
under development contains an unknown concentration of impurities and
has an indefinite compensation. It is a common problem to distinguish
electrostatic fluctuations due to ionized impurities from other causes of disorder \cite{PhysRevB.84.024120, PhysRevB.95.155202}.
Therefore a detailed understanding of DA recombination will be useful
for the characterization of novel semiconductor materials \cite{PhysRevB.84.024120, PhysRevB.95.155202, doi:10.1063/1.4962630, PhysRevB.85.235204}.

In earlier studies DA recombination was analytically described using a model of screened
electrostatic fluctuations which is acceptable in the case of heavily doped
compensated semiconductors. In this approach one uses some phenomenological
parameters like a screening radius, which are not so easy to determine \cite{PhysRevLett.80.2413, doi:10.1063/1.124655, doi:10.1002/pssc.200304243}.
Another method to describe DA recombination is a numerical simulation
based on an algorithm of electrostatic energy minimization. Using this approach
a spectrum of DA photoluminescence line was simulated in the limit of low
pump intensities \cite{PhysRevB.62.8023}. The numerical simulation provides a reasonable
agreement of the calculated spectrum broadening with obtained experimental results.

\begin{figure}
\includegraphics[width=\linewidth]{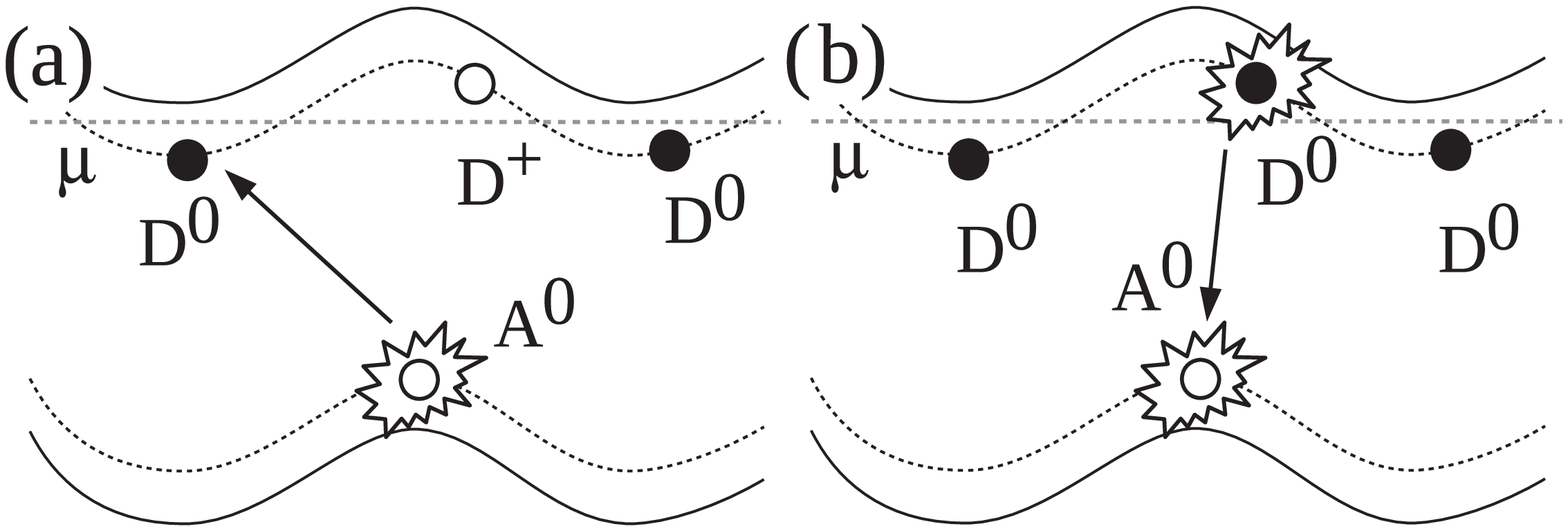}
\caption{\label{fig0} The energy band diagram under different conditions. Photo-excited impurities neutralized via the optical pumping are marked with sparkles.
(a) The case of low pump intensities. Photo-excited, non-equilibrium donors and acceptors are rare and distant therefore
photo-excited, non-equilibrium acceptors recombine with equilibrium donors. (b) The case of high pump intensities.
Photo-excited, non-equilibrium donors and acceptors recombine with each other.
}
\end{figure}

The aim of our study is to generalize the method of numerical simulation for arbitrary pump intensities
and to analyze simulation results using an analytical approach.
In the thermal equilibrium state without optical pumping at $T=0$ all
states of major impurities below the chemical potential $\mu$ are neutral while all states above $\mu$ are ionized.
Here we call such neutral impurities "equilibrium impurities". Under the optical pumping
a part of ionized impurities became neutral. We call such neutral impurities "non-equilibrium impurities".
We will show that the spectrum of DA recombination can be considered as a sum of two different components. 
One of these components corresponds to recombination of non-equilibrium minor impurities
and equilibrium major impurities. Another one is due to recombination of non-equilibrium donors and acceptors with each other.
Corresponding transitions are shown in figure~\ref{fig0}~(a) and figure~\ref{fig0}~(b) respectively.

Hereby we virtually divide all DA transitions into bimolecular and monomolecular contributions. 
In the limit of low compensation we derive two analytical models for the broadening of both DA transition
components taking into account random electrostatic fields of ionized impurities.

It should be pointed out that realistic spectra of DA recombination depend on kinetic parameters such as a generation rate,
a band-to-donor electron capture rate, the probability of the thermal ionization of impurities.
In general, two terms contribute to fluctuations of the optical transition energy: the fluctuations of the initial state energy
and the fluctuations of the final state energy. The equation~\ref{hw} contains the only one fluctuating term ${e^2}/{\varepsilon r}$
due to fluctuations of the distance between ionized impurities in the final state.
On the contrary, in the case of bandgap fluctuations (alloy fluctuations, for instance)
only the energy of the initial state fluctuates. It is easier to take kinetic processes into account
if one considers the only one kind of fluctuations. It was made for the case of DA transition~\cite{PhysRev.140.A202}
as well as for the case of bandgap fluctuation~\cite{doi:10.1002/pssb.2221530222}.
The DA recombination in presence of electrostatic fluctuations depends on both terms of fluctuations.
This makes a simultaneous consideration of the impact of random electric fields and kinetic parameters sufficiently more difficult.
For this reason, in most of our numerical simulations and in the analytical analysis
we neglect any processes besides the radiative recombination. It corresponds to the experimental condition of low temperatures
and fast recombination rates. In trying to qualitatively estimate the influence of kinetics we consider separately
the possibility of energy relaxation processes after the optical excitation in our numerical simulation. 

\section{Numerical simulation}
We numerically simulate the energy distribution of donors and acceptors in compensated
semiconductor at the temperature $T=0$ using the energy minimization algorithm via one-electron hopping \cite{doi:10.1063/1.1699114, PhysRevB.29.4260}.
Every random realization is a cubic volume with periodic boundary conditions containing randomly
distributed donors and acceptors. For definiteness here we discuss the case of an n-type semiconductor.
In our calculations a number of donors equals $n_d =10240$ due to a computer memory limitation. A number of
acceptors is defined as $n_a = K \cdot n_d$ where $K$ is a compensation.
In the paper we present results calculated 
using the optimal donor concentration $N_d a_b^3 = 0.01$ which is close to the metal-insulator transition.
In the case of larger concentrations our approach is not applicable because we consider
the insulator state. Numerical simulations for lower concentrations is an extremely time-consuming process
due to an exponentially small overlap integral in equation (\ref{p}).
For every random realization we find the so-called pseudo-ground energy state using
a procedure described by Shklovskii and Efros in the Chapter 14 of \cite{efros1984electronic}. The procedure
finds out a state with local minimum of energy performing a sequence of one-electron hops which decrease the total energy of the system.
The obtained pseudo-ground state is not a global minimum and the total energy still can be decreased via multi-electron hopping.
For trial we have considered a two-electron hopping in our calculation. However it significantly increased the calculation time
and had no influence on spectra of DA recombination.

Initially all acceptors are charged negatively, $n_a$ donors are charged positively and $n_d-n_a$ donors
are neutral. We find an occupied donor with maximum one-electron potential energy $\epsilon_i$ and
an empty donor with minimum one. If the energy of the occupied state exceeds
the energy of the empty one we transfer the electron and repeat this procedure
for other electrons until it is possible. After every electron transfer we
recalculate all one-electron potential energies $\epsilon_i$ for all impurities.

Then we consider one-electron hops which can minimize the total electrostatic energy of
the DA system. The energy difference of a one-electron hop equals to
$$\Delta_{ij} = \epsilon_j - \epsilon_i -\frac{e^2}{\varepsilon r_{ij}}.$$
We find a pseudo-ground state completing all possible hops with negative $\Delta_{ij}$.

Next we numerically calculate spectra of DA recombination using a following algorithm. In order to simulate an optical pumping of
DA system in the pseudo-ground state we neutralize a part of donors and acceptors and again recalculate all one-electron potential energies.
In a strict sense, the part of neutralized impurities monotonically depends on pump intensity via kinetic parameters of a system.
For simplicity's sake we term the percentage of neutralized donors and acceptors as a pump intensity $I$.
We numerically simulate an energy distribution of the DA transition probability by means of summation of
all transitions between all occupied donors and acceptors using equation~(\ref{p}) and the following formula
$$
\hbar \omega = E_g - E_d - E_a + \frac{e^2}{\varepsilon r_{ij}} + (\epsilon_i - \epsilon_j).
$$

In order to obtain smooth spectrum curves we average data over about 100 thousands of random realizations.
We program our simulation using CUDA parallel computing for the performance improving.
We emphasize that our model has only three independent dimensionless parameters:
a concentration $N_d\cdot a_b^3$, a compensation $K$ and a pump intensity $I$ which
equals to the percentage of photo-excited acceptors.

\begin{figure}
\includegraphics[width=\linewidth]{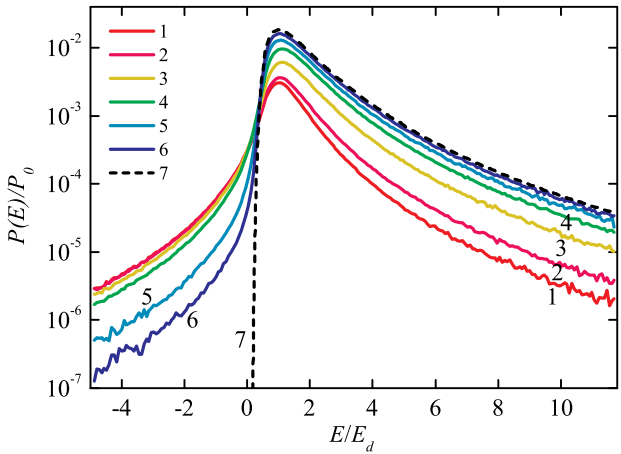}
\caption{\label{fig1} (color online) DA luminescence spectra numerically simulated for different pump intensities. Concentration of donors $N_d a_b^3 = 0.01$, compensation $K = 0.1$.
Pump intensities $I$ are equal to 1 --- 1.56\%, 2 --- 6.25\%, 3 --- 25\%, 4 --- 50\%, 5 --- 85\%, 6 --- 95\% and 7 --- 100\%, $P_0 = 32\pi p_0 N_d a_b^3 /E_d $.
}
\end{figure}

Results of the DA recombination spectra simulation depending on pump intensity are presented in figure~\ref{fig1}.
At low pump intensities the dominating component of spectra is due to recombination of photo-excited acceptors
with equilibrium neutral donors. It is strongly broadened by an electric field of
positively charged donors. This component is located in the low energy side of the spectrum because
these transitions occur at longer distances. While the pump intensity increases the component of photo-excited
carriers grows at the high energy side. An interplay of these components leads to a high energy shift of the
spectrum maxima with increasing pump intensity. Random electric fields broaden the low energy tail.
The high energy tail is due to recombination of closest DA pairs. The influence of random fields on the high energy part of
spectra is weak and one can describe it as $\sim E^{-4}$ similarly to the equation (\ref{FO}).
At high pump intensities the influence of random electric fields decreases due to photo-neutralization of charged impurities.
A simple validation test of our algorithm is to simulate the case of 100\% pump intensity. In this case
all impurities are neutral, random electric fields are absent, and the result coincides
with the equation~(\ref{FO}) as shown in figure~\ref{fig1} with a dashed line.

As mentioned above this algorithm neglects the kinetic behavior of the DA system.
In order to estimate the role of an energy relaxation
we repeat the energy minimization procedure after the photo-excitation and before the spectra calculation.
We considered separately a relaxation of only electrons or only holes.
Results of these simulations will be discussed in section \ref{sec3}.

\section{\label{sec:level2}Analytical calculation of luminescence spectra}
 
In the limit of low compensation all ionized impurities form complexes
due to Coulomb correlation (Chapter 3 of \cite{efros1984electronic}).
Generally an acceptor forms a pair with the
nearest donor. Such pairs are called 1-complexes with respect to the number
of charged donors per one acceptor. Approximately 97.4\% of acceptors form
1-complexes; however some acceptors form 0-complexes and in the vicinity of
some acceptors there are two charged donors which form a 2-complex.

Here we considered the donor-acceptor luminescence of 1-complexes. At low
pumping, an acceptor captures the photo-excited hole from the valence band
and the nearest donor remains charged; therefore a hole recombines with
the next neutral donor. Taking into account the influence of the charged
donor on the transition energy, and neglecting the electric fields of
others, more distant impurities, we can describe the shape of the spectral
line of such transitions analytically. This component of DA recombination is
a monomolecular process and we term this part as a three-center model. 

At high pump intensities, the probability of simultaneous filling of donor
and acceptor from one 1-complex increases. Because of the higher overlap of
the wave functions, the recombination of such pairs dominates in the spectrum.
The influence of electric fields on the transition energy in this case can
be taken into account statistically, using the distribution of the electric field in
a model of randomly located dipoles formed by ionized donors and acceptors.
In this model, we assume that the electric field does not vary at a distance between the donor and the
acceptor which is correct for the limit of small compensation.

In the analytical model we neglect the energy relaxation of charge
carriers on impurities. In a certain approximation, this is equivalent to
the high recombination rate when all trapped carriers recombine
earlier than they are thermally activated back to the band.

We consider a random distribution of photo-excited carriers among DA pairs.
It means that the probability to neutralize an impurity is proportional to $I$.
Therefore the probability to obtain a neutral DA pair is proportional to $I^2$,
while the probability to obtain a partly neutralized DA pair is proportional to $I(1-I)$.
In order to compare spectra with different pump intensities we normalize results by $I$.
It means that spectra in three-center model we multiply by $(1-I)$ while
spectra in random dipoles model have $I$ coefficient.

\subsubsection{The three-center model of the donor-acceptor recombination}

\begin{figure}
\includegraphics[width=4cm]{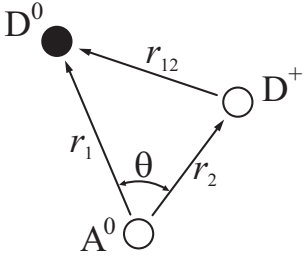}
\caption{\label{fig2} The three-center system consist of two donors and one acceptor.
The donor nearest to the acceptor is ionized and another donor and the acceptor are neutral.}
\end{figure}

At low pump intensities, we consider a three-center system presented in figure \ref{fig2} which consists
of two donors and one acceptor. Initially the three-center system consist of the ionized acceptor,
the ionized donor which is the closest to the acceptor and the neutral equilibrium donor.
Under the condition of weak optical pumping photo-excited holes and
electrons are rare and DA pairs mostly capture only one type of carriers. Consequently, a recombination
mainly occurs between the photo-excited neutral acceptor and the neutral equilibrium donor.
In such a system, the closest to acceptor positively charged donor significantly changes the donor-acceptor
recombination energy, which is equal to

\begin{eqnarray*}
E_{12} =  \frac{e^2}{\varepsilon r_1}+
\frac{e^2}{\varepsilon r_2}-
\frac{e^2}{\varepsilon r_{12}}.
\end{eqnarray*}

where  
\begin{eqnarray*}
r_{12}=\sqrt{r_1^2+r_2^2-2r_1r_2 \cos\theta}.
\end{eqnarray*}

For the three-center system the probability of the photon emission with energy $ E $ can be calculated as

\begin{eqnarray*}
P(E) = p_0 (1-I) \int\limits_0^{+\infty} N_d 4 \pi r_1^2 dr_1 
\int\limits_0^{r_1} N_d 2 \pi r_2^2 dr_2
\int\limits_0^{\pi} \sin\theta d\theta  \nonumber \\
\times \delta (E - E_{12})  \exp  \left( -\frac{2r_1}{a_B} \right)
\exp \left( -\frac{4 \pi}{3} r_2^3 N_d \right)
\end{eqnarray*}

The last factor in the integral describes the probability that the donor at the distance $r_2$
is the closest to the acceptor, and therefore this particular donor is
ionized. In our calculations, this factor is always close to 1 because
$r_1$ and $r_2$ are of the order of the donor radius  $a_B \ll N_d^{-1/3}$.

We proceed from integration over the angle to integration over the transition energy

\begin{eqnarray*}
\frac{\partial E}{\partial \cos \theta} = 
- \frac {e^2 r_1 r_2}{\varepsilon \left( r_1^2+r_2^2-2r_1r_2 \cos\theta \right) ^{3/2}}  \nonumber \\
=  - \left( \frac {\varepsilon}{e^2} \right) ^2 r_1 r_2 \left(  \frac{e^2}{\varepsilon r_1}+
\frac{e^2}{\varepsilon r_2} - E \right) ^3 
\end{eqnarray*}

\begin{eqnarray*}
P(E) = p_0 (1-I) \int\limits_0^{+\infty} N_d 4 \pi r_1^2 dr_1 
\int\limits_0^{r_1} N_d 2 \pi r_2^2 dr_2
\int dE \frac{\partial \cos \theta}{\partial E}  \nonumber \\
\times \delta (E - E_{12})  \exp  \left( -\frac{2r_1}{a_B} \right)
\exp \left( -\frac{4 \pi}{3} r_2^3 N_d \right)
\end{eqnarray*}

The $\delta$ -function allows us to calculate the energy integral

\begin{eqnarray}
\label{three}
P(E) = 8 \pi p_0 (1-I) N_d^2  \left( \frac {e^2}{\varepsilon} \right) ^2 \nonumber \\
\times \int\limits_0^{+\infty} r_1  \exp  \left( -\frac{2r_1}{a_B} \right) dr_1   
\int\limits_0^{r_1} \frac {r_2 \exp \left( -\frac{4 \pi}{3} r_2^3 N_d \right)} 
{ \left( \frac{e^2}{\varepsilon r_1}+
\frac{e^2}{\varepsilon r_2} - E \right) ^3}  dr_2
\end{eqnarray}

Since the further analytical integration is not possible, the calculations were performed numerically under the conditions

\begin{eqnarray*}
E \le  \frac{e^2}{\varepsilon r_1} + \frac{e^2}{\varepsilon r_2} 
- \frac{e^2}{\varepsilon \left( r_1 + r_2 \right)} \\
E \ge  \frac{e^2}{\varepsilon r_1} + \frac{e^2}{\varepsilon r_2} 
- \frac{e^2}{\varepsilon \left( r_1 - r_2 \right)}
\end{eqnarray*}

It is also necessary to exclude from consideration the three-center states in which both donors are charged and form the 2-complex. Such complexes form under the conditions

\begin{eqnarray*}
E - \frac{e^2}{\varepsilon r_1} > \mu \\
E - \frac{e^2}{\varepsilon r_2} > \mu.
\end{eqnarray*}

Here $\mu$ is a chemical potential.
In the limit of low compensation the chemical potential can be found analytically  $\mu = 0.99 N_d^{1/3} e^2 / \varepsilon$(Chapter 3 of \cite{efros1984electronic}).
There is no analytical expression for chemical potential in the case of an intermediate compensation but it can
be calculated numerically \cite{0022-3719-12-10-018}.

\begin{figure}
\includegraphics[width=\linewidth]{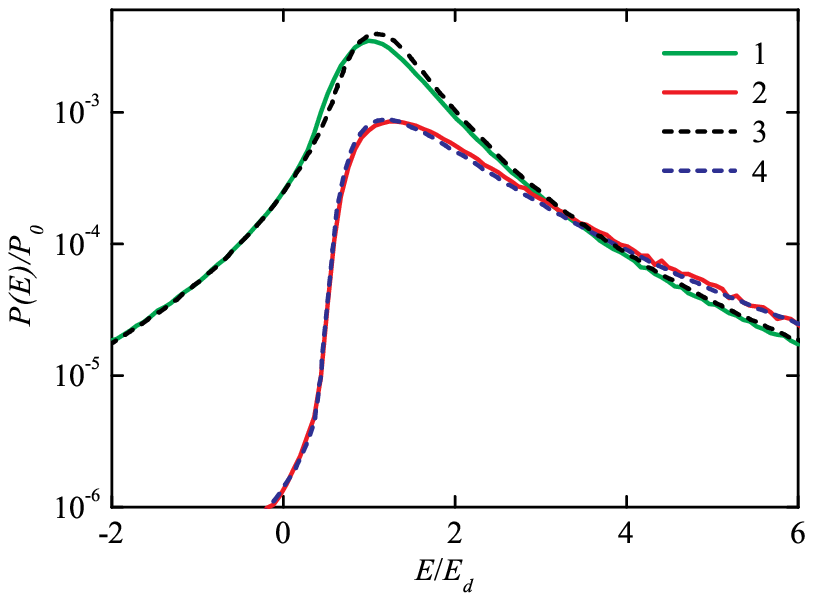}
\caption{\label{fig3} (color online) Luminescence spectra of equilibrium (curves 1 and 3) and photo-excited (curves 2 and 4) electrons.
Dashed lines show spectra calculated analytically using equations (\ref{three}) and (\ref{fields}).
Solid lines are obtained by numerical simulations. Concentration of donors $N_d a_b^3 = 0.01$, compensation $K = 0.025$, pumping $I = 6.25\%$,
$P_0 = 32\pi p_0 N_d a_b^3 /E_d $.}
\end{figure}

\begin{figure}
\includegraphics[width=\linewidth]{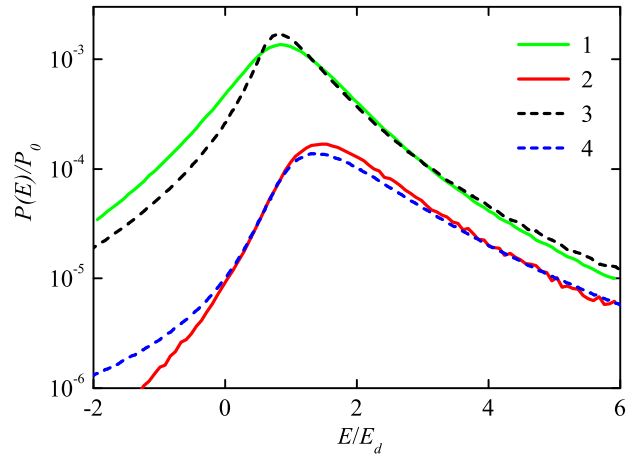}
\caption{\label{fig4} (color online) Luminescence spectra of equilibrium (curves 1 and 3) and photo-excited (curves 2 and 4) electrons.
Dashed lines show spectra calculated analytically using equations (\ref{three}) and (\ref{fields}).
Solid lines are obtained by numerical simulations. Concentration of donors $N_d a_b^3 = 0.01$, compensation $K = 0.5$, pumping $I = 6.25\%$,
$P_0 = 32\pi p_0 N_d a_b^3 /E_d $.}
\end{figure}

Numerically simulated spectra of transitions between photo-excited acceptors and equilibrium neutral donors and
luminescence spectra calculated in the three-center model are shown in figures \ref{fig3} and \ref{fig4} (curves 1 and 3, correspondingly).
At low compensation (figure \ref{fig3}) our analytical calculations
are in a good agreement with our simulation results. At high compensation (figure \ref{fig4}) a difference between two
curves arises. The three-center model underestimates the broadening of the low-energy tail,
however the qualitative similarity still persists. 


\subsubsection{Random dipoles model}

At high pumping, the main contribution to the luminescence spectrum is made by the recombination of
photo-exited acceptors and donors from one 1-complex. The donor and acceptor within one pair
are close to each other, so the transition probability is much higher than
for the centers from different pairs. The main line broadening for closely
located states is due to random electric fields in the material. The
transition energy in an electric field $F$ can be written as

\begin{eqnarray*}
E = \frac{e^2}{\varepsilon r} + eFr \cos \theta 
\end{eqnarray*}

Here $\theta$ is the angle between the direction of the electric field and
the line connecting the donor and the acceptor. This formula is valid if
the electric field is constant on the scale of the recombination length,
which is well satisfied only for low compensation. In the opposite case of the high compensation
the recombination length is comparable with the mean distance between ionized impurities,
therefore our approach is not applicable.

An electrically neutral system of randomly distributed point charges can be
represented as a set of randomly oriented dipoles.
Such an approach for dipoles with equal dipole moment modulus was first
used by Holtsmark \cite{Holtsmark} to describe electric fields in plasma.
Shklovskii and co-authors considered a more
general case when the magnitude of the dipole moment modulus is also random
for the photoconductivity of compensated semiconductors \cite{kogan1980electric}. In this case
the distribution function of a random electric field modulus $F$  is given
by

\begin{eqnarray*}
f(F) = \frac{4}{\pi F_{md}} \frac{(F/F_{md})^2}{ \left[ 1 + (F/F_{md})^2 \right] ^2} 
\end{eqnarray*}

Here $F_{md}$ is the most probable electric field. 

\begin{eqnarray*}
F_{md} = 2.515 \frac {KeN_d^{2/3}}{\varepsilon}
\end{eqnarray*}

Then the probability of emission of a photon with an energy $E$ can be calculated as

\begin{eqnarray*}
P(E) = p_0 I \int\limits_0^{+\infty} N_d 2 \pi r^2 dr   \exp \left( -\frac{2r}{a_B} -\frac{4 \pi}{3} r^3 N_d \right) \nonumber \\
\times\int\limits_0^{\pi} \sin \theta d \theta 
\int\limits_0^{+\infty} dF \delta  \left( E - \left(  \frac{e^2}{\varepsilon r} + eFr\cos \theta \right) \right) f(F)
\end{eqnarray*}

For further calculations we will make a change of the variable in the integral over the electric field to $T = F \cos \theta $. 
We will separately consider the integral over $\theta $ in which we will change the variable to $x = \cos \theta$ and denote $ T/F_{md} = \alpha $

\begin{eqnarray*}
2 \int\limits_0^{\pi /2} \sin \theta d \theta \frac {1}{\cos \theta}
\frac {\left(  \cfrac{T}{F_{md} \cos \theta} \right) ^2} 
{\left[ 1+ \left( \cfrac{T}{F_{md} \cos \theta} \right) ^2 \right] ^2} = \nonumber \\
= 2 \int\limits_0^{1} \frac {\left( \alpha / x \right) ^2}  
{\left[ 1+ \left( \alpha / x \right) ^2 \right] ^2}
\frac {dx}{x} = \frac{1}{1+\alpha^2}
\end{eqnarray*}

Then the equation for the probability of transition can be rewritten as

\begin{eqnarray*}
P(E) = 4 p_0 I N_d \int\limits_0^{+\infty} r^2 dr \exp \left( -\frac{2r}{a_B} -\frac{4 \pi}{3} r^3 N_d \right) \nonumber \\
\times\int\limits_{-\infty}^{+\infty} dT  \delta \left( E - \left(  \frac{e^2}{\varepsilon r} - eTr \right) \right) 
 \cfrac{F_{md}}{F_{md}^2 + T^2}
\end{eqnarray*}

after integration over $T$, we obtain the final equation which was calculated numerically

\begin{eqnarray}
\label{fields}
P(E) = {4 p_0 I N_d} \int\limits_0^{+\infty} r dr \exp \left( -\frac{2r}{a_B} -\frac{4 \pi}{3} r^3 N_d \right) \nonumber \\
\times \cfrac{e F_{md} r^2}{e^2 F_{md}^2 r^2 + \left( E - \cfrac{e^2}{\varepsilon r} \right)^2}
\end{eqnarray}

Let us note that in the limiting case of 100\% pumping, all the impurity
centers are neutral and do not produce random Coulomb fields. In this case
$F_{md} \rightarrow~0$ and the fraction $\cfrac{1}{\pi} \cfrac{e F_{md} r}{e^2 F_{md}^2 r^2 + \left( E - \cfrac{e^2}{\varepsilon r} \right)^2}$ 
converges to $\delta \left( E - \cfrac{e^2}{\varepsilon r}
\right)$.
In this case the resulting formula~\ref{fields} goes over into equation (\ref{FO}) which was obtained without taking into account random fields.

In figures \ref{fig3} and \ref{fig4} the numerically simulated spectra of transitions between photo-excited donors and acceptors
are compared with the results of analytical calculation in the random dipoles model (curves 2 and 4, correspondingly).
In the case of low compensation, a good agreement of the results
is observed. There is a discrepancy between the analytical calculation and the simulation results at the low-energy tail at high compensation.
Transitions with low energies correspond to recombination of very distant DA pairs. For such DA pairs
a magnitude of the electric field is not constant over the size of the pair. As a result the
random dipoles model for such low transition energies is not applicable because it overestimates
the line broadening.

\section{\label{sec3}Discussion}

Figures \ref{fig3} and \ref{fig4} show the DA recombination spectra for an
n-type semiconductor with a donor concentration $N_d a_b^3 = 0.01$ and
a pump intensity $I = 6.25\%$ at different compensation levels, calculated both by numerical
simulation and analytically. The relaxation of photo-excited
carriers was not taken into account here.

In the numerical simulation of the photoluminescence spectrum, we
separately calculated the luminescence of equilibrium and photo-excited
donors. As was said in the theoretical part, the first case is well
described by the three-center model, and the second one corresponds to
random dipoles.

Such a two-component description allows us to explain a complicated behavior of DA recombination
at different experimental conditions. Monomolecular and bimolecular
components have different energy positions and its interplay
leads to a high energy shift of DA line with increasing pump intensity.
This energy shift is one of the known features of DA recombination and
was observed experimentally~\cite{Zacks}. In the limit of high pump intensity
the energy of the DA transition coincides with the energy of a free electron to acceptor recombination.
In this case these two lines overlap and could be resolved only by different behavior in a magnetic field.

Monomolecular component dominates at low pump intensities and its
electrostatic broadening does not depend on pump intensity.
The broadening of a bimolecular component decreases with the pump intensity
which is observed experimentally at high pump intensities \cite{doi:10.1063/1.4825317}.

\begin{figure}
\includegraphics[width=\linewidth]{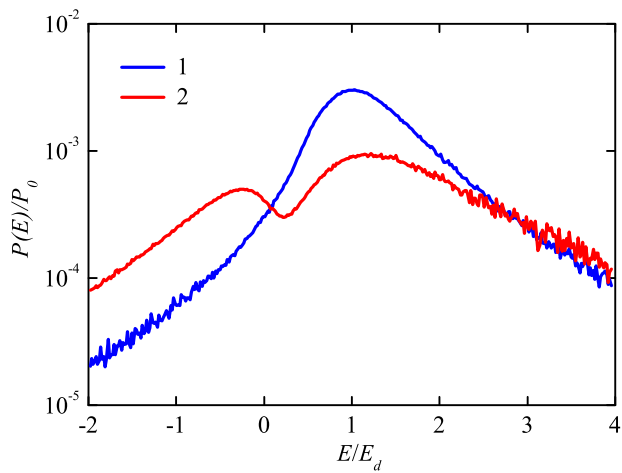}
\caption{\label{fig5} (color online) Luminescence spectra of DA recombination before (curve 1) and after (curve 2) full relaxation of photo-excited minority carriers
Concentration of donors $N_d a_b^3 = 0.01$, compensation $K = 0.1$, pumping $I = 1\%$, $P_0 = 32\pi p_0 N_d a_b^3 /E_d $. }
\end{figure}

In the case of a fast recombination rate when relaxation processes could be neglected
two components of DA recombination are strongly overlapped and can not
be spectrally resolved. However this situation can change if the energy relaxation rate is comparable
with the recombination rate.
We obtain the most demonstrative results of our numerical simulation
in the case when minority photo-excited carriers relax before recombination
while photo-excited majority carriers recombine without
relaxation. This case is close to the experimental condition of p-type
semiconductor because trapped electrons relax faster than holes.
In figure~\ref{fig5} we present results of such simulation:
two components of DA recombination are clearly resolved.

Two separate DA lines are often present in low temperature luminescence spectra
of semiconductors \cite{doi:10.1063/1.328788, PhysRevLett.25.1614}. A usual interpretation of this result involves a presence of two
different kinds of impurity in a sample. However, our analysis shows that such two components,
especially close ones, can originate from one kind of impurity.

\section{Conclusion}

The donor-acceptor luminescence spectra in compensated semiconductors were
calculated numerically and analytically. The spectral and kinetic
properties of luminescence at various pump intensities, concentrations, and
compensations were studied. It was shown that the
donor-acceptor recombination line consists of two components, the
recombination of photo-excited holes with photo-excited electrons and the
recombination of photo-excited holes with equilibrium electrons. These two
components have different broadening mechanisms due to Coulomb correlations
and random electric fields. The spectral form of both 
components in the low compensation limit have been obtained
analytically. Numerical simulation results show that these simple
analytical models not only agree with the numerical simulation at low
compensation, but also qualitatively describe the spectra up to
moderate compensations. Taking into account
the possibility of photo-excited carriers relaxation we show that
the two components can be resolved in the DA recombination spectra.

\begin{acknowledgments}
We acknowledge funding from Russian Foundation of Basic Researches (project No 17-02-00539).
This research was partly supported by Presidium of Russian Academy of Science:
program No 8 "Condensed matter physics and new generation of materials". N.S.A. thanks
the Foundation for the Advancement of Theoretical Physics and Mathematics "BASIS".
\end{acknowledgments}

\end{document}